\documentclass[twoside]{article}

\renewcommand{\thefootnote}{\fnsymbol{footnote}}  

\font\bfit=cmbxti10 at 10pt

\catcode`\@=11
\long\def\@makefntext#1{
\protect\noindent \hbox to 3.2pt {\footnotesize\hskip-.9pt  
$^{{\@thefnmark}}$\hfil}#1\hfill}                  

\def\thefootnote{\fnsymbol{footnote}}
\def\@makefnmark{\hbox to 0pt{$^{\@thefnmark}$\hss}}        

\def\fnt#1#2{\footnotetext{\kern-.3em
          {$^{\mbox{\scriptsize #1}}$}{#2}}}

\renewcommand{\thefootnote}{\fnsymbol{footnote}}
\newcommand{\alphfootnote}
          {\setcounter{footnote}{0}
           \renewcommand{\thefootnote}{\scriptsize\alph{footnote}}}

\renewcommand{\thefootnote}{\fnsymbol{footnote}}  

\def\ps@myheadings{\let\@mkboth\@gobbletwo
\def\@oddhead{\hbox{}
\rightmark\hfil\footnotesize\thepage}   
\def\@oddfoot{}\def\@evenhead{\footnotesize\thepage\hfil
\leftmark\hbox{}}\def\@evenfoot{}
\def\sectionmark##1{}\def\subsectionmark##1{}}

\def\runninghead#1#2{\pagestyle{myheadings}
\markboth{{\protect\footnotesize\it{\quad #1}}\hfill}
{\hfill{\protect\footnotesize\it{#2\quad}}}}
\def\fpage#1{\begingroup
\voffset=.3in
\thispagestyle{empty}\begin{table}[b]\centerline{\footnotesize #1}
          \end{table}\endgroup}

\oddsidemargin=\evensidemargin
\renewcommand\section{\@startsection {section}{1}{\z@}%
                                  {-10.5pt \@plus -1pt \@minus -.2pt}%
                                  {3.5pt \@plus.2ex}
                                   {\bf}}
\renewcommand\subsection{\@startsection{subsection}{2}{\z@}%
                                    {-8.5pt\@plus -1pt \@minus -.2pt}%
                                    {2.5pt \@plus .2pt}
                                     {\bfit}}
\renewcommand\subsubsection{\@startsection{subsubsection}{3}{\z@}%
                                     {-8.5pt\@plus -1pt \@minus -.2pt}%
                                     {2.5pt \@plus .2pt}%
                                     {\it}}

\newcounter{appendixc}
\newcounter{subappendixc}[appendixc]
\newcounter{subsubappendixc}[subappendixc]
\renewcommand{\thesubappendixc}{\Alph{appendixc}.\arabic{subappendixc}}
\renewcommand{\thesubsubappendixc}
          {\Alph{appendixc}.\arabic{subappendixc}.\arabic{subsubappendixc}}

\renewcommand{\appendix}[1] {\vspace{12pt}
        \refstepcounter{appendixc}
        \setcounter{figure}{0}
        \setcounter{table}{0}
        \setcounter{lemma}{0}
        \setcounter{theorem}{0}
        \setcounter{corollary}{0}
        \setcounter{definition}{0}
        \setcounter{equation}{0}
        \renewcommand{\thefigure}{\Alph{appendixc}.\arabic{figure}}
        \renewcommand{\thetable}{\Alph{appendixc}.\arabic{table}}
        \renewcommand{\theappendixc}{\Alph{appendixc}}
        \renewcommand{\thelemma}{\Alph{appendixc}.\arabic{lemma}}
        \renewcommand{\thetheorem}{\Alph{appendixc}.\arabic{theorem}}
        \renewcommand{\thedefinition}{\Alph{appendixc}.\arabic{definition}}
        \renewcommand{\thecorollary}{\Alph{appendixc}.\arabic{corollary}}
        \renewcommand{\theequation}{\Alph{appendixc}.\arabic{equation}}
        \noindent{\bf Appendix \theappendixc #1}\par\vspace{5pt}}
\newcommand{\subappendix}[1] {\vspace{12pt}
        \refstepcounter{subappendixc}
        \noindent{\bf Appendix \thesubappendixc. {\kern1pt \bfit #1}}
          \par\vspace{5pt}}
\newcommand{\subsubappendix}[1] {\vspace{12pt}
        \refstepcounter{subsubappendixc}
        \noindent{\rm Appendix \thesubsubappendixc. {\kern1pt \it #1}}
          \par\vspace{5pt}}

\newcommand{\textlineskip}{\baselineskip=13pt}
\newcommand{\smalllineskip}{\baselineskip=10pt}

\newcommand{\copyrightheading}[4]
  {\vspace*{-2.5cm}\smalllineskip{\flushleft
  {\footnotesize Quantum Information and Computation, {Vol.~#1}, {No.~#2}
 {(#3)} {#4}}\\
  {\footnotesize \copyright\kern2pt Rinton Press}\\
    }}
\def\abstracts#1#2#3{{
          \centering{\begin{minipage}{4.5in}\footnotesize\baselineskip=10pt
          \parindent=0pt #1\par 
          \parindent=15pt #2\par
          \parindent=15pt #3
          \end{minipage}}\par}}

\renewenvironment{thebibliography}[1]
        {\frenchspacing
         \small\rm\baselineskip=11pt
         \begin{list}{\arabic{enumi}.}
        {\usecounter{enumi}\setlength{\parsep}{0pt}     
          \setlength{\leftmargin 12.7pt}{\rightmargin 0pt}
         \setlength{\itemsep}{0pt} \settowidth
          {\labelwidth}{#1.}\sloppy}}{\end{list}}

\newcounter{itemlistc}
\newcounter{romanlistc}
\newcounter{alphlistc}
\newcounter{arabiclistc}

\newcommand{\fcaption}[1]{
        \refstepcounter{figure}
        \setbox\@tempboxa = \hbox{\footnotesize Fig.~\thefigure. #1}
        \ifdim \wd\@tempboxa > 5in
           {\begin{center}
        \parbox{5in}{\footnotesize\smalllineskip Fig.~\thefigure. #1}
            \end{center}}
        \else
             {\begin{center}
             {\footnotesize Fig.~\thefigure. #1}
              \end{center}}
        \fi}

\newcommand{\tcaption}[1]{
        \refstepcounter{table}
        \setbox\@tempboxa = \hbox{\footnotesize Table~\thetable. #1}
        \ifdim \wd\@tempboxa > 5in
           {\begin{center}
        \parbox{5in}{\footnotesize\smalllineskip Table~\thetable. #1}
            \end{center}}
        \else
             {\begin{center}
             {\footnotesize Table~\thetable. #1}
              \end{center}}
        \fi}

\def\pmb#1{\setbox0=\hbox{#1}
          \kern-.025em\copy0\kern-\wd0
          \kern.05em\copy0\kern-\wd0
          \kern-.025em\raise.0433em\box0}

\def\FigName{figure}%
\newbox\captionbox
\long\def\@makecaption#1#2{%
  \ifx\FigName\@captype
    \vskip\abovecaptionskip
    \setbox\tempbox\hbox{{\figurecaptionfont #1\hskip1em #2}}
          \ifdim\wd\tempbox< 28pc
          \centerline{\box\tempbox}
          \else
          {\figurecaptionfont #1\hskip1em #2\par}
\fi\else
          \setbox\tempbox\hbox{{\tablecaptionfont #1\hskip1em #2}}
          \ifdim\wd\tempbox< 28pc 
          \centerline{\box\tempbox}
          \else
          {\tablecaptionfont #1\hskip1em #2\par}%
          \fi   
 \vskip\belowcaptionskip
 \fi}
\InputIfFileExists{psfig.sty}
{\typeout{^^Jpsfig.sty inputed...ok}}{\typeout{^^JWarning: psfig.sty could be be found.^^J}}
\InputIfFileExists{epsfsafe.tex}
{\typeout{^^Jepsfsafe.tex inputed...ok}}
                              {\typeout{^^JWarning: epsfsafe.tex could not be found.^^J}}
\InputIfFileExists{epsfig.sty}
{\typeout{^^Jepsfig.sty inputed...ok}}{\typeout{^^JWarning: epsfig.sty could not be found.^^J}}
\InputIfFileExists{epsf.sty}
{\typeout{^^Jepsf.sty inputed...ok}}{\typeout{^^JWarning: epsf.sty could not be found.^^J}}%
\def\fps@figure{tbp}
\def\ftype@figure{1}
\def\ext@figure{lof}
\def\fnum@figure{Fig.\ \thefigure}
  \usepackage{amsfonts}
  \usepackage{amssymb}
 \usepackage{mathrsfs}
 \usepackage{amsmath}
 \usepackage{color}
 \usepackage{soul}
 \usepackage{dsfont}
 \usepackage{ifpdf}
  \ifpdf
 \usepackage{epstopdf}   
 \fi

\newcommand{\ket}[1]{\ensuremath{\vert#1\rangle}}

 \newcommand{\id}{\mathds{1}}

\newcommand{\tr}{\ensuremath{\mathrm{tr}}}

\def\ket#1{\left|#1\right>}

\begin{document}
\setlength{\textheight}{8.0truein}    

\runninghead{Majorana fermions and non-locality}
          {E. T. Campbell, M. J. Hoban and J. Eisert}

 \normalsize\textlineskip
 \thispagestyle{empty}
 \setcounter{page}{1}
 
\vspace*{-2.5cm}

\vspace*{0.88truein}

 \alphfootnote

 \fpage{1}

\vspace*{0.035truein}
\centerline{\bf MAJORANA FERMIONS AND NON-LOCALITY}
\vspace*{0.37truein}
\centerline{\footnotesize
EARL T.\ CAMPBELL\footnote{earltcampbell@gmail.com.}}
\vspace*{0.015truein}
\centerline{\footnotesize\it Dahlem Center for Complex Quantum Systems, Freie Universit{\"a}t Berlin,}
\baselineskip=10pt
\centerline{\footnotesize\it Berlin, Germany }
\vspace*{10pt}
\centerline{\footnotesize 
MATTY J.\ HOBAN}
\vspace*{0.015truein}
\centerline{\footnotesize\it ICFO-Institut de Ci\`{e}ncies Fot\`{o}niques, Mediterranean Technology Park,}
\baselineskip=10pt
\centerline{\footnotesize\it Castelldefels (Barcelona), Spain}
\vspace*{10PT}
\centerline{\footnotesize JENS EISERT}
\vspace*{0.015truein}
\centerline{\footnotesize\it Dahlem Center for Complex Quantum Systems, Freie Universit{\"a}t Berlin,}
\baselineskip=10pt
\centerline{\footnotesize\it Berlin, Germany }
\vspace*{0.225truein}

\vspace*{0.21truein}

\abstracts{Localized Majorana fermions emerge in many topologically ordered systems and exhibit exchange statistics of Ising anyons. This enables noise-resistant implementation of a  limited set of operations by braiding and fusing Majorana fermions. Unfortunately, these operations are incapable of implementing universal quantum computation.  We show that, regardless of these limitations,  Majorana fermions could be used to demonstrate non-locality (correlations incompatible with a local hidden variable theory) in experiments using only topologically protected operations.  We also demonstrate that our proposal is optimal in terms of resources, with 10 Majorana fermions shown to be both necessary and sufficient for demonstrating bipartite non-locality.   Furthermore, we identify severe restrictions on the possibility of tripartite non-locality. We comment on the potential of such entangled systems to be used in quantum information protocols.}{}{}

\vspace*{10pt}

\vspace*{3pt}

\vspace*{1pt} \textlineskip  

\section{Introduction}
\noindent
Fermions that are their own anti-particle are known as Majorana, as opposed to Dirac, fermions.  While presently there is no evidence that any fundamental particles are Majorana fermions, they frequently emerge as localised quasi-particles in models of condensed matter systems~\cite{Kitaev01,FuKane08,Alicea10,Lutchyn10,Oreg10}. Recent years have seen a race to experimentally confirm their existence, with some evidence already found~\cite{Mourik12,Das12,Deng12,Rokhinson12,Churchill13,Finck13}.  Research into these systems is  driven, at least partially, by their potential applications in topological quantum computing.  For localised and well-separated Majorana fermions with zero energy, the system is protected from noise effects that could otherwise prove devastating in quantum computers.  

Suitable Majorana fermions emerge in many two-dimensional (2D) systems including the Kitaev honeycomb lattice~\cite{Kitaev2006}, fractional quantum hall systems~\cite{Moore91,Nayak96}, topological insulators~\cite{FuKane08,FuKane09}, and a variety of other systems~\cite{Alicea12}.  Adiabatically exchanging these fermions, and so braiding their world-lines, gives rise to the non-abelian exchange statistics of Ising anyons.  Majorana fermions can also emerge as edge modes in one-dimensional (1D) systems~\cite{Alicea12}, such as the Kiteav wire~\cite{Kitaev01}.  While braiding is not a meaningful concept in strict 1D systems, networks of wires also enable braiding with Ising statistics~\cite{Alicea11}.   The unitary evolution from braiding is geometric in origin, and so robust against small experimental imperfections. However, these topologically protected braiding operations are not computationally powerful enough for universal quantum computing.  Indeed, any free fermionic system can be efficiently classically simulated~\cite{Bravyi05b,Piotr}.  Access to some non-topological operations --- which may be noisy, but not too noisy~\cite{VirPlen,Howard11,Howard12b,deMelo13} --- can be used to promote the system to full universality~\cite{BraKit05,Bravyi06,Meier13,Bravyi12,Jones12}.  However, here we are interested in understanding the purely topologically protected capabilities of Majorana fermions, and in particular their capacity for demonstrating non-locality \cite{WernerWolf}.    

Non-locality is the inability for a local hidden variable (LHV) theory to reproduce the correlations of space-like separated measurements. Until now, non-locality has only been investigated in more exotic topological systems by Brennen \emph{et al.}~\cite{Brennen09}.  Having only considered systems capable of universal quantum computation, Brennen \emph{et al.} concluded their work saying, ``\emph{it is intriguing to ask whether one could find intermediate anyonic theories which have the power to generate Bell violating states by topologically protected gates, but are not universal for topological quantum computation}".  We resolve this mystery by showing that Majorana fermions, and equivalently Ising anyons, could be used in an experiment demonstrating the non-locality of quantum mechanics.  It appears that the standard and ubiquitous Clauser-Horne-Shimony-Holt (CHSH) inequality cannot be violated with only topological operations, and the non-topological resources sufficient for a CHSH violation have been investigated~\cite{Howard12}.  We turn instead to a non-local experiment proposed by Cabello~\cite{Cabello01a,Cabello01} where 2 parties each select from 3 possible measurements, with each measurement producing 3 bits of classical information as outcomes.   This non-locality proposal was built on the idea of the Mermin-Peres ``magic square", which was originally used to show the contextuality of quantum mechanics~\cite{Mermin90b,Peres90}.   We find that a variant of these experiments could be implemented with each of two parties holding 5 Majorana fermions, and present a Bell inequality based on the magic square.  We also present no-go results showing for two parties holding 4 Majoranas each, all experimental statistics can be produced by a local hidden variable theory.  If two parties unequally share Majoranas, say Alice holds $n$ and Bob holds $m$ with $n<m$, we find that the resource is locally equivalent to both parties holding just $n$ Majorana fermions.  Hence, 10 Majorana fermions are
both necessary and sufficient for the phenomenon of bipartite non-locality to be topologically demonstrated.  Furthermore, we find that the correlations required to demonstrate the three-party Greenberger-Horne-Zeilinger (GHZ) paradox~\cite{Mermin90,Anders09,Hoban11} cannot be implemented with any number of Majoranas.   This indicates that our proposal is the simplest possible non-locality experiment with Majorana fermions.  

Bell experiments were intended, initially, to falsify alternative theories that claim the world to be local in Nature.  Now these Bell-type non-locality experiments are also known to have practical applications.  Such experiments can form the basis of quantum cryptography when the devices used are faulty or even untrusted~\cite{Acin07,Barrett12}.  Typically, such proposals are envisaged for photonic systems, since they provide an easier means of accomplishing space-like separation of measurement events (a requirement for a non-local experiment). Polarisation-entangled photons have, for many years, been used to test Bell's proposal, proving successful if one ignores the detector loophole~\cite{Aspect82}.  Experiments with trapped ions have efficient enough measurements to avoid the detector loophole, but do not satisfy space-like measurement separation~\cite{Rowe01}.  Designing better experiments, hopefully closing all loopholes, is an active area of research~\cite{Buhrman03,Brunner12}.  We shall not argue that our topological proposal is more promising than the aforementioned approaches.  Rather, we take the first step by showing theoretical feasibility under ideal conditions.  Implicit in our approach is the assumption that the physical system can be shared between two parties over sufficient distances that measurements are space-like separated; this is a similar technical difficulty faced by ion trap designs~\cite{Rowe01}.

\section{Majorana fermions and braiding}
\noindent
\subsection{Majorana fermions}
\noindent
We begin with a review of Majorana fermions and their available dynamics, largely following 
Refs.~\cite{Bravyi06,Alicea12}. Generally, Majorana fermions are described by a set of Hermitian operators $ c_1,\dots,  c_{2n}$ satisfying 
\begin{equation}
	\{ c_j, c_k\} = 2\delta_{j,k}
\end{equation}
and $ c_j^\dagger =  c_j$ for all $j$, acting on the physical Hilbert space ${\cal H}= {\cal H}_0\oplus {\cal H}_1$ constituting a direct sum between the even and odd parity sectors.  The algebra of physical operators ${\cal F} = {\cal L}({\cal H})$ is
spanned by products of an even number of Majorana fermion operators.
Taking two such fermions and clockwise braiding their world lines results in a unitary 
$U({j,k})$
%
that maps the operators as 
\begin{equation}
	U({j,k})  c_a U({j,k})^\dagger =
	\left\{
	\begin{array}{ll}
	 c_a & {\rm  if }\, a\neq \{j,k\},\\
	 c_k & {\rm if }\, a =j,\\
	-  c_j & {\rm if }\, a =k,
	\end{array}
	\right.
\end{equation}
for $k>j$.
Composing these braid operations results in a permutation $P\in S_{2n}$, with possible phase change $Q$, so that 
${c}_{j}\mapsto (-1)^{Q_{j}} {c}_{P_{j}}$.  The phases are constrained so that the global parity is preserved, 
leaving $\prod_{j=1}^{2n} {c}_{j}$ unchanged. 

Such braidings are a special case of unitary transformations $U$ acting on ${\cal H}$ that reflect linear mode transformation 
\begin{equation}\label{TF}
	 c_j\mapsto  U c_j' U^{\dagger}  = \sum_{k=1}^{2n} V_{j,k} c_k
\end{equation}
for $V\in SO(2n)$.
Such unitary transformations are the ones commonly 
considered in the context of fermionic linear optics.
All states encountered in this work are Gaussian fermionic states \cite{Bravyi05b,Piotr,Kastoryano}. They are
entirely described by their anti-symmetric covariance matrix $\gamma\in \mathbb{R}^{2n\times 2n}$, $\gamma=-\gamma^T$,
which has entries
$\gamma_{j,k} = i{\rm tr}(\rho[ c_j, c_k])/2$.

\subsection{Stabiliser language}
\noindent
We primarily describe quantum states in the Heisenberg picture by specifying a sufficient number of eigenvalue equations.  
We say an operator $s\in {\cal F}$ stabilises a state vector $\ket{\psi}$ if 
\begin{equation}
	s \ket{\psi} = \ket{\psi}. 
\end{equation}	
We assume that initialisation of the system prepares a state vector $\ket{\psi_{0}}$ stabilised by $g_{j}=i {c}_{2j-1}{c}_{2j}$ for all $j\in\{1,\dots,n\}$. 
%
%
Therefore, any state prepared from the initialisation $\ket{\psi_{0}}$ by braiding will be stabilised by $g_{j}=\pm i {c}_{P_{2j-1}}{c}_{P_{2j}}$ for some 
permutation $P\in S_{2n}$.
Note that, since permutations are one-to-one mappings, if $j\neq k$, then $P_{j}\neq P_{k}$.  Products of stabilisers are again stabilisers, and so $g_{j}$ generate a group $\mathcal{S}$ associated with the state.  Similarly, topologically protected measurements, also called charge measurements, are those of the form $i{c}_{j}{c}_{k}$.  Throughout, we say a state is \textit{accessible} if and only if it can be prepared with the above described topological operations. Again, states obtained by performing such measurements on 
Gaussian states are Gaussian.

Collective charge measurements, such as of ${c}_{1}{c}_{2}{c}_{3}{c}_{4}$, cannot be measured non-destructively and in a topological manner.  The collective charge observable can, however, be inferred by measuring $i{c}_{1}{c}_{2}$ and $i{c}_{3}{c}_{4}$ and multiplying the outcomes.  However, such a process is destructive in that it will always disentangle these Majoranas from all other systems.  We labour this point because the capacity to make non-destructive charge measurements increases the computational power of the system~\cite{Bravyi06} beyond that assumed in our later no-go theorems.   

\subsection{Anyonic formalism}
\noindent
We shall restate some of the above in the anyonic formalism, which may be of benefit to some readers.  Using, $\sigma$ to denote an Ising anyon, $\psi$ for another fermion, and $\id$ for the vacuum, we have the fusion channel $\sigma \times \sigma \rightarrow \id + \Psi$.  Fusing Ising anyons $j$ and $k$ is equivalent to measuring $i c_{j}c_{k}$, where producing a $\Psi$ particle is equivalent to the $-1$ measurement outcome (eigenvalue) and producing the vacuum $\id$ outcome denotes the $+1$ eigenvalue.  Conversely, if we begin with a vacuum and create $n$ pairs of Ising anyons, we have the initialisation state described above if anyons from the $j^{\mathrm{th}}$ pair are labeled as $2j-1$ and $2j$.

\subsection{Entanglement properties}
\noindent
We proceed by identifying the equivalence classes of entangled states that are accessible under the operations described above.   Consider two parties, Alice and Bob, holding respective sets of Majorana fermions $A= \{1,2,\dots,n \}$ and $B=\{n+1,\dots, n+m\}$.  Clearly, any state stabilised by $\pm i{c}_{j}{c}_{k}$ can be locally prepared by Alice for $j,k \in A$ labeling a pair of Majorana fermions.  A similar statement holds for Bob, and so the interesting stabilisers have  $j \in A$ and $k \in B$ with each party holding one half of a pair of Majorana fermions  (henceforth referred to as Majorana pairs).  Assuming they hold $N$ such pairs, they can always locally braid such that the state is stabilised by $i{c}_{j}{c}_{j+N}$ for all $1 \leq j \leq N$.   We see that the entanglement is entirely captured by the number of such Majorana pairs shared by Alice and Bob, and throughout we assume this canonical form for the stabilisers.    For notational simplicity, it is beneficial to use ${a}_{j}={c}_{j}$ and ${b}_{j}={c}_{j+n}$, so Majorana pairs are stabilised by $i {a}_{j} {b}_{j}$.   

Crucially, the number of Majorana pairs is distinct from the number of Bell pairs that give rise to useful entanglement.  To investigate the useful correlations between Alice and Bob, we must consider how such pairs respond to measurements.  Consider a pair of stabilisers $i{a}_{j}{b}_{j}$ and $i{a}_{k}{b}_{k}$ for $j\neq k$, 
the state is also an eigenstate of their product $(i {a}_{j} {b}_{j})(i {a}_{k} {b}_{k})$ which using anti-commutation of fermions equals $-(i {a}_{j} {a}_{k})(i {b}_{j} {b}_{k})$.  The factors $i {a}_{j} {a}_{k}$ and $i {b}_{j} {b}_{k}$ are locally measurable, and so their outcomes must be anti-correlated.  Hence, the state has the flavor of a singlet state.

To make measurements Alice and Bob must share at least two Majorana pairs, but two alone is trivial since there is only one possible local measurement and with only one measurement we cannot construct a test of non-locality.  With three Majorana pairs,  Alice has three possible measurement observables that mutually anti-commute and are isomorphic 
to the Pauli operators. An encoding is a collection of maps ${\cal F}\rightarrow {\cal B}({\cal H})$, identifying the set of fermionic modes with qubit or
spin systems.  The identification of products of Majorana fermions with Pauli operators of an associated qubit system can be taken as
\begin{eqnarray}
	X=i{a}_{1}{a}_{2} ,      Y=i{a}_{1}{a}_{3} , Z=i{a}_{2}{a}_{3}.
\end{eqnarray}
From the anti-commutation relations of the Majorana fermions, one can readily verify that these operators generate the Pauli group of a single qubit.
Bob similarly has measurement options isomorphic to the Pauli  operators. The correlations between Alice and Bob resulting from the above measurements on three Majorana pairs match those of Pauli  measurements on the two-qubit singlet state (for Alice and Bob each having a single qubit).  Hence, three Majorana pairs can be said to reproduce the entanglement of a Bell pair.  However, it is well-known that the measurement statistics under the operations allowed here can be reproduced by a local hidden variable (LHV) theory.  Later we present a strengthened proof that even \emph{four} Majorana pairs are incapable of demonstrating non-locality.  

\section{Non-locality}
\noindent
Before proceeding we refine the concepts of non-locality.  Assume Alice and Bob hold some quantum state $\rho$ and are able to freely choose from a set of measurements $\{ \mathcal{A}_{j} :j\in I\}$ and $\{ \mathcal{B}_{k}: k \in I \}$ respectively, where $I$ denotes the set of different kinds of measurement settings. 
The measurements will have possible outcomes that we denote $\alpha,\beta\in R$ for a suitable $r$ that occur with probability
\begin{equation}
    P(\alpha, \beta | j , k) =  \tr [ (  \Pi_{j,\alpha} \otimes \Pi_{k,\beta} ) \rho ]
\end{equation}
where $\Pi_{j,\alpha} \geq 0 $ ($\Pi_{k,\beta}\geq 0$) is the positive-operator valued measure for setting $j$ ($k$) and outcome $\alpha$ ($\beta$).  There are many different experiments that can achieve the same probability distributions, and each of these is a realisation of $P$.  Conversely, we say a probability distribution $P$ is quantum if there exists a choice of measurements and a quantum state $\rho$ that realises $P$.

When are these observations a proof of non-locality? Imagine that Alice and Bob are space-like separated and are allowed to make their choice of measurement freely, they now want to know whether any classical model can produce these observations. Since they cannot communicate, this classical model has some pre-determined instructions that, given Alice and Bob's choice of measurements, will output some value.  Such an instruction set is called a LHV theory.  An arbitrary hidden variable $\lambda$ takes values in some space $\Lambda$ equipped with a probability measure, and which determines local probability functions
\begin{equation}
	\lambda  \mapsto  p_A( {\alpha} |j,\lambda) ,
	\,
	\lambda  \mapsto  p_B( {\beta} | k,\lambda).
\end{equation}
The probability of obtaining the outcome pair $(\alpha,\beta)$, given that $j,k$ have been chosen by Alice and Bob, is then
\begin{equation}
    P( {\alpha}, {\beta} | j, k) =  \int dM(\lambda) p_A( {\alpha} |j,\lambda)  p_B( {\beta} | k,\lambda).
\end{equation}
A probability distribution is local (also called a LHV) if and only if there exists a decomposition of the above form, and we denote this as $P \in \mathcal {P}_{L}$.  If no such local model exists, $P \notin \mathcal {P}_{L}$, we say the probability distribution is non-local.  Furthermore, any experiment realizing a non-local probability distribution is a non-locality experiment.  

Typically, we confirm non-locality by checking whether a probability distribution violates a Bell inequality.  Let us fix the number of measurement settings and outcomes, and denote the entire set of possible probability distributions by $\mathcal {P}$.  The Bell inequalities follow by first defining a real-valued linear function, $G: \mathcal {P} \rightarrow \mathbb{R}$, which is sometimes called a non-local game~\cite{Brassard,Toner,Cabello01a}, the respective parties are referred to as players and the 
action taken are strategies. The non-local game, as the term is used here, is described by a real-valued function, 
$V: R \times R \times I\times I\rightarrow \mathbb{R}$ such that
\begin{equation}
\label{non-localGame}
    G(P) = \sum_{\alpha, \beta, j, k} V(\alpha, \beta, j, k) P(\alpha, \beta | j, k).
\end{equation}
For any such non-local game, there exists a Bell inequality that holds for all local $P$
\begin{equation}
\label{generalBell}
    G(P) \leq \Omega_{c}(G) := \sup \{ G(P) : P \in \mathcal {P}_{L} \} .
\end{equation}
The above is true simply by definition.  However, for any $P$ and $G$ where $G(P)>\Omega_{c}(G)$ we can conclude $P$ is non-local.  

Let us consider a particular non-local game $G$, and its classical limit $\Omega_{c}(G)$.  Since non-local games are linear functions and $\mathcal {P}_{L}$ is a convex set, the classical limit can always be achieved by an extremal point in $\mathcal {P}_{L}$.  That is, the value of $\Omega_c(G)$ can always be achieved by a probability distribution of the form
\begin{equation}
    P( {\alpha}, {\beta} | j, k) =   c_A( {\alpha} |j )  c_B( {\beta} | k ),
\end{equation}
where Alice and Bob deterministically assign measurement outcomes.  There are only finitely many such distributions, 
so this greatly simplifies the evaluation of $\Omega_{c}(G)$.

\section{Magic squares with Majoranas}
\noindent
To demonstrate that these Majorana fermions exhibit non-locality we adapt a specific proof of ``non-locality without inequalities" first devised by Cabello~\cite{Cabello01a,Cabello01}. There exist other proofs of non-locality for qubits without the use of a Bell inequality, such as those by Hardy~\cite{Hardy92} and the 
GHZ paradox~\cite{Mermin90,Anders09,Hoban11}.
Crucially, Hardy's argument does not work for maximally entangled states and so it cannot be applied. Furthermore we show later that the GHZ  argument does not apply here either. Cabello's proof works for two Bell states, a four-qubit state, shared by Alice and Bob each of which can make a measurement on two qubits of this state. The analysis shares much of its character with the proof that quantum mechanics is non-contextual derived by Peres and Mermin, often referred to as the ``Magic Square Game". We will follow the simplification of Cabello's approach as presented by Aravind~\cite{Aravind02} so that the result may be of more general appeal, but optimally tailored to systems of Majorana fermions.

\begin{table}
 \tcaption{Squares describing measurements made by Alice and Bob on their respective parts of their quantum state (made up of two Bell pairs with one half of each pair held by each party).  Measuring these operators on five Majorana pairs, Alice and Bob yield identical measurement outcomes for the entry they have in common in their respective squares.  Notice that, up to a phase accounting for correlation vs. anti-correlation, the tables are equivalent under interchanging ${a}_{j}$ with ${b}_{j}$.  
 Furthermore, for each column the product of the three entries yields $+\id$, whereas for each row the product of all entries gives $-\id$.}
\centering
	\begin{tabular}{c}

\begin{tabular}{r|c|c|c|}

\multicolumn{1}{c}{} & \multicolumn{3}{c}{{Alice}} \\

\multicolumn{1}{c}{} & \multicolumn{1}{c}{$\mathcal{A}_{1}$} &  \multicolumn{1}{c}{$\mathcal{A}_{2}$}  & \multicolumn{1}{c}{$\mathcal{A}_{3}$} \\ \cline{2-4}
&	$-a_1 a_3 a_4 a_5 $ 	  & $ i a_1 a_4 				$ & $ i a_3 a_5$ 			\\ \cline{2-4}
&	$i a_3 a_4 $ 			  & $- a_1 a_2 a_3 a_4 	   	$ & $-i a_{1}a_{2}$ 		\\ \cline{2-4}
&	$i a_1 a_5 $ 			  & $ i a_2 a_3				$ & $ a_1 a_2 a_3 a_5$  \\ \cline{2-4}
	\multicolumn{1}{c}{} & \multicolumn{3}{c}{} \\ 
	\multicolumn{1}{c}{} & \multicolumn{3}{c}{{Bob}} \\ 	\cline{2-4}
	
	$\mathcal{B}_{1}$    &$-b_1 b_3 b_4 b_5 $ 	  & $- i b_1 b_4 				$ & $- i b_3 b_5$ 			\\ \cline{2-4}
	$\mathcal{B}_{2}$    &	$-i b_3 b_4 $ 			  & $-b_1 b_2 b_3 b_4 	   	$ & $i b_{1}b_{2}$ 		\\ \cline{2-4}
	$\mathcal{B}_{3}$    &	$-i b_1 b_5 $ 			  & $ -i b_2 b_3				$ & $ b_1 b_2 b_3 b_5$  \\ \cline{2-4}
	\end{tabular} 
	\end{tabular}
\label{MagicSquare}
\end{table}

\subsection{The set-up}
\noindent
We now describe the set-up. We assume that Alice and Bob share five Majorana pairs, prepared in the canonical form outlined earlier. 
Each party then makes a choice of three different measurement settings.
Each measurement setting relates to three measurements involving a subset of the Majorana modes held by the respective party. Hence,
for each measurement setting, a string of three values of $\pm 1$ is obtained, so $R=\{\pm1\}^{\times 3}$. 
For Alice we label the choice of measurement as 
$\mathcal{A}_{j}=(\mathcal{A}_{1,j},\mathcal{A}_{2,j}, \mathcal{A}_{3,j})$
for $j\in\{1,2,3\}= I$ and for Bob the choice of measurement is written as $\mathcal{B}_{k}= (\mathcal{B}_{k,1},\mathcal{B}_{k,2}, \mathcal{B}_{3,k}$)
for $k\in I$.  Alice and Bob then obtain the output strings 
${\alpha}=(\alpha_{1}, \alpha_{2}, \alpha_{3})$, 
${\beta}=(\beta_{1}, \beta_{2}, \beta_{3})\in \{\pm 1\}^{\times 3}$, respectively, with $j,k\in I$ labeling the choice of measurements $\mathcal{A}_{j}$ and $\mathcal{B}_{k}$.  The measurements are fixed by the columns (for Alice) and rows (for Bob), of square tables (so-called ``magic squares"), as shown in Table~\ref{MagicSquare}.  In these magic squares, Alice and Bob's respective square tables contain, up-to a phase, the same measurements performed by the respective parties. It can easily be verified that an element of Alice's table multiplied with the same element of Bob's table (i.e. the same $j^{\mathrm{th}}$ row and $k^{\mathrm{th}}$ column in each table) gives a stabiliser of their shared quantum state. Formally, this enforces that $\alpha_{k}=\beta_{j}$ for all $j,k\in I$.  Furthermore, the product of observables in any column of both tables gives $+\id$, and so for any measurement setting $\alpha_{1} \alpha_{2} \alpha_{3}=1$.  Whereas the product of observables in any row of both tables gives $-\id$, and so $\beta_{1} \beta_{2} \beta_{3}=-1$.   Notice that any triple of observables, for either Alice or Bob, always contains a single measurement observable acting on 4 Majorana modes.  As remarked earlier, such measurements cannot be directly measured but can be inferred from other measurement outcomes.  Here this poses no problem as the required information is provided by the remaining pair of observables.  For instance, the first column for Alice corresponds to measuring $\{ -a_1 a_3 a_4 a_5,  i a_3 a_4 , i a_1 a_5   \}$, and while $ -a_1 a_3 a_4 a_5$ cannot be directly measured, we see that $( i a_3 a_4) (  i a_1 a_5 ) =  -a_1 a_3 a_4 a_5$ and so we can simply infer the outcome from $\alpha_{1}=\alpha_{2}\alpha_{3}$.  In any actual experiment, Alice and Bob only measure a pair of observables, but for clarity of exposition it is convenient to speak of each party measuring a whole triple of observables.

Aside from these constraints, the measurement outcomes are entirely random and so the experiment realises
\begin{equation}
\label{majoranaProb}
    P^{*}({\alpha}, {\beta} | j, k) = \bigg\{ \begin{array}{l}
    \frac{1}{8} \mbox{ if } ( \alpha_{k}=\beta_{j} ) \wedge ( \alpha_{1}\alpha_{2}=\alpha_{3} ) \wedge ( \beta_{1}\beta_{2}=-\beta_{3} ), \\ 
    0 \mbox{ otherwise.}
     \end{array} 
\end{equation}
and in the next section we confirm this to be non-local and robust against some experimental noise.

\subsection{The magic square game without inequalities}
\noindent
Earlier, we saw that for any non-local game the best local strategy can be achieved by Alice and Bob deterministically assigning measurement outcomes.  In the problem at hand, it is convenient to describe these local probability distributions by a $3\times 3$ table with entries in $\{\pm 1\}$. 
We use $T^{A}$ to denote Alice's table, which has entries such that
\begin{equation}
    c_{A}({\alpha}|j) = \left\{\begin{array}{l} 1  \mbox{ if } (T_{1,j}^{A},T_{2,j}^{A},T_{3,j}^{A}) = {\alpha}, \\
    0 \mbox{ otherwise}.
 \end{array} 
 \right.
\end{equation}
Whereas for Bob, we take
\begin{equation}
    c_{B}({\beta}|k) = \left\{
    \begin{array}{l} 1  \mbox{ if } (T_{k,1}^{B},T_{k,2}^{B},T_{k,2}^{B}) = {\beta}, \\
    0 \mbox{ otherwise}.
 \end{array} 
 \right.
\end{equation}
Clearly, there is a one-to-one correspondence between these tables and deterministic probability distributions.  Note also the slight difference in definitions for Alice and Bob. For Alice, a single measurement setting will output a column of her table, whereas for Bob, a single measurement setting will output a row of his table.  This convention is justified because the quantum correlations satisfy $\alpha_{k}=\beta_{j}$ for all $j,k\in I$, and so if the LHV theory replicates this correlation we have that 
\begin{equation}
	T^{A}_{k,j}=T^{B}_{k,j}. 
\end{equation}
 Therefore, Alice and Bob must share identical tables to reproduce this feature of the  quantum correlations.  However, the quantum correlations have an additional feature, they satisfy parity constraints. Alice's parity constraint entails that for all $j$, we have $\prod_{l\in I} T^{A}_{l,j}=1$, and hence the parity of the entire table is $\prod_{l,j\in I} T^{A}_{l,j}=1$.  In contrast, Bob's parity constraint entails that for all $k$, we have $\prod_{l\in I} T^{B}_{k,l}=-1$, and so $\prod_{l,k\in I} T^{B}_{k,l}=-1$.  We conclude that Alice and Bob cannot hold identical classical tables and also satisfy all the parity constraints, and so they can never perfectly reproduce the quantum correlations.   Since quantum observables do not commute, finding the parity of the table of operators can change depending on the order we multiply the entries. 

\subsection{The magic square game with Bell inequalities}
\noindent
We have seen that classical experiments can never reproduce the quantum correlations demonstrated in the previous section.  However, we are interested in what level of imperfection can be tolerated by any quantum experiment while still violating locality.  To quantify this we must specify a particular non-local game, see Eq.~(\ref{non-localGame}), 
by specifying a function $V$.  The goal is to have a larger value when the correlations match those of the quantum predications, and smaller when they fail.  The simplest choice is a function that takes two values, and it is conventional to take these as $\pm 1$, and so we have
\begin{equation}
    V({\alpha}, {\beta}, j, k) = \left\{ \begin{array}{l}
    1 \mbox{ if } ( \alpha_{k}=\beta_{j} ) \wedge ( \alpha_{1}\alpha_{2}=\alpha_{3} ) \wedge ( \beta_{1}\beta_{2}=-\beta_{3} ), \\ 
    -1 \mbox{ otherwise.}
     \end{array}
     \right.
\end{equation}
This non-local game we call the magic square game.  It is easy to verify that using Majorana fermions, which realise the probability distribution $P^{*}$ (see Eq.~\ref{majoranaProb}), we find $G(P^{*})=9$. 

We are now in a position to identify $\Omega_c(G)$, the maximum classical value.  Alice and Bob could use identical tables, but they then contravene many parity constraints, and we find this results in at most $G(P)=3$.  However, the maximum is achieved if Alice and Bob use classical tables that differ only in a single entry and satisfy all the parity constraints, such as in Table~\ref{classical}.  This yields $G(P)=7=\Omega_{c}(G)$ since for 8 of the measurement settings we acquire a contribution of $1$, but for one setting we have $-1$. Hence, any imperfect experiment with Majorana fermions that attains $G(P)>7$ is sufficient to demonstrate non-locality.  

 \begin{table}
	\centering
	\tcaption{An attempted classical strategy for the magic square game, where Alice and Bob satisfy all of their row/column constraints.  To achieve this they must differ in at least 1 entry, which here is the bottom right entry.  When the referee specifies the bottom row for Bob and right-most column for Alice, this strategy will fail.  However, it will win in the other 8 choices of row and columns.}

\begin{tabular}{ccc}

{Alice} & & {Bob} \\


 \begin{tabular}{|c|c|c|}
	\hline
	+1 & +1 & -1 \\ \hline
	+1 & -1 & +1 \\ \hline
	+1 & -1 & \emph{-1} \\ \hline
	\end{tabular}
	 & &
	 \begin{tabular}{|c|c|c|}
	\hline
	+1 & +1 & -1 \\ \hline
	+1 & -1 & +1 \\ \hline
	+1 & -1 & \emph{+1}  \\ \hline
	\end{tabular}
	\end{tabular}
\label{classical}
\end{table}

We also comment on how the imperfect quantum setting can, sometimes, economically be described.  For perfect operations of the type considered here, the covariance matrix $\gamma$ will only contain entries
contained in $\{0,-1,1\}$ and satisfies $\gamma^2=-\id$.
If errors and imperfections are present, the entries of $\gamma$ will also attain values different from those, 
while it is still true that 
\begin{equation}
	-\gamma^2\leq \id.
\end{equation}	
Under braiding transformations of the form (\ref{TF}), covariance matrices transform as congruences
$\gamma\mapsto V\gamma V^T$ with orthogonal matrices $V$, 
which can easily be kept track of. The statistics of measurements can still be determined based on the covariance 
matrix only, even if errors are taken into account in the preparation step.

\section{Too few Majorana pairs}

\subsection{Setting}
\noindent
Earlier we saw that if Alice and Bob share three Majorana pairs, the available measurements are isomorphic to Pauli  measurements on a single Bell pair.  It is well-known folklore that such a system can be modeled by a local-hidden variable theory.    Here we present a stronger argument covering up to 4 Majorana pairs.   This shows that five or more Majorana pairs are necessary, as well as sufficient, to demonstrate non-locality.  We begin by characterizing the full set of possible measurements.  For Alice, there are three pairs of commuting measurements she can perform
\begin{eqnarray}  
	\mathcal{A}_{1} &=& ( i a_{1}a_{2} , i a_{3}a_{4} ) , \\ 
	\mathcal{A}_{2} &=& ( i a_{1}a_{3} , i a_{4}a_{2} ) , \\
	\mathcal{A}_{3} &=& ( i a_{1}a_{4} , i a_{2}a_{3} ) .
\end{eqnarray}  
If Alice measures a pair of observables $\mathcal{A}_{j}$, she get two random bits, which we denote $\alpha_{1}, \alpha_{2} \in \{ +1, -1  \}$.  Similar measurement sets, labeled $\mathcal{B}_{k}$, are available to Bob, with $b_{j}$ replacing $a_{j}$. If Bob measures the analogous set, so $k=j$, he gets perfectly anticorrelated outcomes, such that $\beta_{1}=-\alpha_{1}$ and $\beta_{2}'=-\alpha_{2}'$.   Next we consider when Bob measures a different set of operators, so $k \neq j$.  We need to determine which products of measurements correspond to stabilisers of the 4 Majorana pairs.   We observe that for Alice and any measurement set, the product of the observables is $-a_{1}a_{2}a_{3}a_{4}$, and similarly for Bob the product is $-b_{1}b_{2}b_{3}b_{4}$.   Since the observables are matching and contain an even number of fermionic operators, these observables will have correlated outcomes.   Formally, these correlations entail that $\alpha_{1} \alpha_{2} = \beta_{1} \beta_{2}$  for all measurement settings $j$ and $k$.  However, when $j \neq k$, this is the only correlation and otherwise the measurement outcomes are entirely random.  From these observations we can deduce that the outcome probabilities, as a function of measurement settings $j$ and $k$, are
\begin{equation}
\label{Pconditions}
	P({\alpha}, {\beta} | j, k) =  \left\{ \begin{array}{ll}   \frac{1}{4} 
& \mbox{if }   (j=k) \wedge (\alpha_{1}=-\beta_{1}) \wedge ( \alpha_{2}=-\beta_{2} )   ,\\
0 & \mbox{if } ( j=k) \wedge  ( \alpha_{1} \neq-\beta_{1} \mbox{ or } \alpha_{2} \neq-\beta_{2} ),  \\
\frac{1}{8}  & \mbox{if } (j \neq k ) \wedge ( \alpha_{1} \alpha_{2}= \beta_{1} \beta_{2} ) ,\\
0  & \mbox{if } ( j \neq k)  \wedge ( \alpha_{1} \alpha_{2} \neq \beta_{1} \beta_{2} ).
  \end{array}\right.
\end{equation}
The structure is richer than for Pauli  measurements on a Bell pair.  However, it can still be explained by an LHV theory, which we now turn to.  

\subsection{Local hidden variable model}
\noindent
For our LHV theory we use a set of four hidden variables, which we label as $\lambda= \{ \nu_{1}, \nu_{2}, \nu_{3}, \mu \}$, each of which takes values $\{ +1, -1\}$.  
Hence, we can write 
\begin{equation}
	\int M(d\lambda) = \sum_\lambda p(\lambda).
\end{equation}
We distinguish $\mu$ from the other variables as it plays a unique role.  Next we fix Alice's probabilities to depend deterministically on the hidden variables as follows
\begin{eqnarray}
	p_{A}({\alpha} | j, \lambda) = 
	\left\{\begin{array}{ll}	1 & \mbox{if } \alpha_{1}=\nu_{j}, \alpha_{2}=\mu \nu_{j} ,\\
	0 & \mbox{otherwise},
	\end{array}
	\right.
\end{eqnarray}
and for Bob we take
\begin{eqnarray}
	p_{B}({\beta} | k, \lambda) = 
	\left\{
	\begin{array}{ll}	1 & \mbox{if } \beta_{1}=-\nu_{k}, \beta_{2}=-\mu \nu_{k} , \\
	0 & \mbox{ otherwise}.
	\end{array}
		\right.
\end{eqnarray}
Notice that, for all measurement settings, the measurement outcomes obey 
\begin{equation}	
	\alpha_{1} \alpha_{2} = \mu,\,\beta_{1} \beta_{2} = \mu. 
\end{equation}
Hence, for all choices of the hidden variables
$\lambda$, the measurements outcomes satisfy $\alpha_{1} \alpha_{2} = \beta_{1} \beta_{2}$.  Furthermore, when $j=k$ we have the added constraint that measurements satisfy $\alpha_{1}=-\beta_{1}$ and $\alpha_{2}=-\beta_{2}$.  This tells us that for all choices of hidden variables, the distributions satisfy the second and fourth lines of Eqs.~(\ref{Pconditions}).  To achieve the correct weighting of the non-zero probabilities we simply take a uniform distribution over the hidden variables so that $p(\lambda) =1/16$ for all $\mu$ and $\nu_{j}$.  This completes our account of a LHV theory for all possible measurements on 4 Majorana pairs.

\section{Impossibility of GHZ state preparation}
\label{NOghz}
\noindent
Here we show that the correlations of 3-party GHZ states cannot be prepared using topologically protected operations and Majorana fermions, blocking attempts to violate locality using the Mermin-GHZ paradox.  The proof technique can be easily extended to larger GHZ states, and the associated generalizations of the GHZ paradox~\cite{Hoban11}.  We begin with some definitions and general observations.   We consider products of fermionic operators, of the form
\begin{equation}
	S = i^{q} \prod c_{j}^{w_{j}} ,
\end{equation}
where $w_{j} \in \{ 0,1 \}$ and $q \in \{0, 1,2,3\}$.  For two such operators, $S$ and $S'$, denote with
$S \star S'$ the overlap in fermionic operators they hold in common, so that  $S \star S' = \sum_{j} w_{j} w_{j}'$.  Furthermore, let $S \star S' \star S''$ denote the overlap shared by all three operators, which is $S \star S' \star S''=  \sum_{j} w_{j}w_{j}'w_{j}''$.  We will show that, for any accessible state with stabiliser $\mathcal{S}$, both the following hold
\begin{enumerate}
	\item[(i)] for all $S,S' \in \mathcal{S}$, we have that $S \star S'$ is even; 
	\item[(ii)] for all $S,S', S'' \in \mathcal{S}$, we have that $S \star S' \star S''$ is even.
\end{enumerate}
Accessible states possess a stabiliser $\mathcal{S}$ generated by $g_{j}= \pm i c_{P_{2j-1}}c_{P_{2j}}$, where $P\in S_{2n}$. 
The structure of the generators imposes a structure on the whole group, which we use to prove the above.  First, note that distinct generators $g_{j}$ share no fermionic operators in common, and so $g_{j} \star g_{k} = 2 \delta_{j,k}$.  Expressing the stabilisers of an accessible state in terms of the generators, we have
\begin{equation}
	S = \prod_{k} g_{k}^{u_{k}} , 
\end{equation} 
for some $u_{k}\in \{0,1\}$. 
For two such operators, each generator they share contributes a pair of fermionic operators in common, and so
\begin{equation}
	S \star S' = 2 \sum_{k} u_{k} u_{k}' .
\end{equation} 
 Clearly, this is always even.  Furthermore,
 \begin{equation}
 	S \star S' \star S''  = 2 \sum_{k} u_{k} u_{k}' u_{k}'' ,
\end{equation} 
which is again even.  More generally, overlaps between larger collections of operators must, for accessible states, again be even.  Note also that, property (i) also follows from the necessary commutativity of the group $\mathcal{S}$, and is true for all quantum states.  Whereas property (ii) is a genuine constraint on the whole set of quantum states. 

The correlations required for the Mermin paradox can be achieved by measuring Pauli operators on a GHZ state stabilised by
\begin{eqnarray}
 S&=&	X_{1}Z_{2}Z_{3}  , \\  
 S'&=&	Z_{1}X_{2}Z_{3}  , \\  
 S''&=&Z_{1}Z_{2}X_{3} , \\ 
 S'''&=&-X_{1}X_{2}X_{3} ,
\end{eqnarray}
using some encoding ${\cal F}\rightarrow {\cal B}({\cal H})$.
From the anti-commutation of the Pauli  operators, that is $X_{j}Z_{j}=-Z_{j}X_{j}$, we see that the last stabiliser follows from the first three, such that $S'''=S''S'S$.   We consider all possibilities where the Pauli  operators are replaced by appropriate products of Majorana operators.  To be measurable, every $X_{j}$ and $Z_{j}$ must consist of a product of an even number of Majorana operators.  As remarked earlier, it is essential $X_{j}$ anti-commutes with $Z_{j}$ and so $X_{j} \star Z_{j} = \gamma_{j}$ where $\gamma_{j}$ is odd.  Finally, they must be local operators, so operators associated with different parties must be supported on distinct subsets of Majorana modes.  From locality we deduce that
\begin{eqnarray}
	S \star S' \star S'' &=& (X_{1} \star Z_{1})+(X_{2} \star Z_{2})+(X_{3} \star Z_{3})  \nonumber  \\
	&=& \gamma_1 + \gamma_2 + \gamma_3 .
\end{eqnarray}
From anti-commutivity, all of the $\gamma_{j}$ are odd, and the sum of 3 odd numbers is again odd.   We conclude that such a set of stabilisers contradicts property (ii) of accessible states.  Under very general assumptions we have shown that no accessible states are isomorphic to the 3-qubit GHZ state.  In particular, the most natural encoding would be to take $Z_{j}=i c_{3j}c_{3j+1}$ and $X_{j}=i c_{3j}c_{3j+2}$, for which $\gamma_{j}=1$ and $S' \star S'' \star S'''=3$.  However, more exotic encoding beyond this more canonical choice are also covered by the above argument.  Alternative proofs based on the fermionic Gaussian nature of the states generated by topological operations are conceivable, and could potentially make use of Wick's theorem.

\section{Quantum information protocols with Majorana fermions}
\noindent
We finally comment on the perspective of using Majorana fermions in basic protocols of quantum information processing involving entanglement. Surely, all of the consequences of braiding operations can be classically efficiently simulated, by virtue of the observation that they constitute a subset of those operations in fermionic linear optics. Still, a number of interesting quantum information protocols involving entanglement can readily be conceived which are sketched here. 
This analysis complements the findings of Ref.\ \cite{Freedman}, in which measurement-only topological quantum computation has been considered.

\subsection{Teleportation}
\noindent
Notably, instances of {\it teleportation} \cite{BBCJPW01a}
are possible. Meaningful variants of teleportation involving Majorana fermions
should share the features that (i) an unknown state is considered the input, taken from a set of at least two non-orthogonal quantum
states, and (ii) the output should be statistically indistinguishable from the input to the protocol.

Here, Alice and Bob not only share $n$ Majorana pairs, so as usual, the initial state is stabilised by $i a_{j}b_{j}$ for  $ j=1,\dots, n$. But furthermore, Alice holds further 
$n$ Majorana modes that are not entangled with Bob. These local ancilla are stabilised by 
\begin{equation}
	i a_{j}a_{j+1},\,\, j={n+1, n+3,\dots , 2n-1}. 
\end{equation}
and represent the modes that will be teleported, herein called the input modes. Such a scheme can indeed be devised by making use of 4 Majorana pairs, so $n=4$.  The input register consists of 4 Majorana modes, initially stabilised by $ia_5 a_6$ and $ia_7 a_8$.   Many different inputs could be prepared by braiding the input modes.  For instance, the input can be set to one of two non-orthogonal inputs by either (I) braiding $a_6$ and $a_7$  or (II) not. It is not difficult to see that these two situations (I) and (II) cannot be perfectly distinguished with unit probability.

The remainder of the protocol does not depend on this input, and neither is knowledge of it required. In the next step, 
$i a_5 a_1$, 
$i a_6 a_2$, 
$i a_7 a_3$, 
$i a_8 a_4$ 
are measured and the results, elements of $\{\pm 1\}^{\times 4}$, 
classically communicated to Bob. One can then show that 
the output statistics of measurements performed by Bob, $i b_1 b_2$ and $i b_3 b_4$, appropriately interpreted using the received classical bits, is indistinguishable from 
the respective measurements on $i a_5 a_6$ and $i a_7 a_8$. For example, in scenario (II), if the outcomes $(-1,-1,-1,-1)$ are communicated, the 
outcomes of measurements $i b_1 b_2$ and $i b_3 b_4$| are completely determined to be $ +1$.

\subsection{Dense coding}
\noindent
In a similar fashion, {\it dense coding} \cite{Dense}
can be performed with Majorana fermions and the above specified operations. Again, let us be specific what a fair analogue of such a scheme would be. 
A valid dense coding scheme is one in which the entanglement-assisted single shot classical capacity of a quantum channel is higher compared to 
the corresponding capacity in the absence of entanglement. We hence would like to introduce a protocol
where starting from entangled shared resources, 
with local braiding operations and a subsequent transmission of some modes, one can encode more bits of 
classical information than is possible without having shared entangled resources available,
but transmitting the same number of modes. For qudits, it is known that the entanglement-assisted capacity is exactly double the one 
reachable without
assistance \cite{WernerDenseCoding}. For Majorana fermions, it turns out, 
the same holds true.

This is readily possible with 3 Majorana pairs, again partially shared by Alice and Bob and partially held by Alice. 
Both parties share 2 Majorana pairs, stabilised by
$i a_1 b_1$, 
$i a_2 b_2$. In addition, Alice holds 2 Majorana modes in a state stabilised by $i a_3 a_4$. By local braiding, Alice can achieve a state
stabilised by
\begin{equation}
	 \biggl\{i (-1)^{\gamma_1} a_1 b_1,
	i (-1)^{\gamma_2} a_2 b_2,
	i (-1)^{\gamma_1+\gamma_2} a_3 a_4:  \gamma_1,\gamma_2\in \{0,1\}
	\biggr\},
\end{equation}
in a scheme that maps 
\begin{equation}
	a_j \mapsto (-1)^{\gamma_j} a_j, 
\end{equation}
$\gamma_j\in\{0,1\}$ for $j=1,\dots, 4$, and fixing $\gamma_3=\gamma_1+\gamma_2$ and $\gamma_4=1$.
It is easy to see that with a suitable choice of $\gamma_1,\gamma_2\in\{0,1\}$, 2 bits of information can be encoded.  Furthermore,  after Alice transmits Majorana modes $a_{1}$ and $a_{2}$ to Bob, he can reliably retrieve $\gamma_1$ and $\gamma_2$ by locally measuring $ia_{1}b_{1}$ and $ia_{2}b_{2}$.  At the same time, with two Majorana modes and no shared entanglement, a single bit of information can be encoded only. This constitutes a 
valid dense coding scheme based on Majorana fermions. It also resembles the situation of transmitting two bits of classical information with a single
transmitted qubit, if entangled resources are initially available.

\section{Closing remarks}
\noindent
We have seen that locality can be violated using only the topologically protected operations of Majorana fermions or Ising anyons.  These operations are a subclass of fermionic linear optics~\cite{Bravyi05b}, and so we conclude that these systems can also violate locality.  The model of fermionic linear optics was proposed as an analog of bosonic linear optics, where a well-known LHV theory prevents any non-locality experiment~\cite{Bell}. Seen in this light, our result is quite surprising. Furthermore, we remark that that fractionalized analogs of Majorana fermions, known as {\it parafermions}~\cite{Fendley12}, may face similar obstacles from the existence of certain hidden variable theories~\cite{Gross06,Veitch12,Mari12}. It is natural to ask more generally what other anyonic systems are non-local, again without the requirement of universal quantum computing capabilities.  
We know of only one related study~\cite{Wootton}, where a six-level classical spin model was used to simulate the charge submodel of $D(S_{3})$ (the quantum double model based on the symmetric group of 3 objects). Although non-locality was not explicitly discussed, it is clear that the classical spin model amounts to a LHV theory.  Finally, we speculate that these experiments could prove useful as a probe to certify the presence of Majorana fermions and potentially help dispel some of the present ambiguity surrounding experiments.

In the final stages of completing this research we became aware of Ref.~\cite{Deng13}, which addresses to some extent similar questions and concerns the non-locality of Majorana fermions when nondestructive collective-charge measurements are available.   This represents an additional experimental capability, posing additional challenges, beyond braiding and standard measurements.  However, without this capability, the GHZ correlations required by Ref.~\cite{Deng13} cannot be implemented by virtue of our no-go result of Sec.~\ref{NOghz}.

\section{Acknowledgements}
\noindent
This work has been supported by the BMBF (QuOReP), the EU (Q-ESSENCE,SIQS), the ERC (TAQ), the EURYI, the EPSRC and the CHIST-ERA (DIQIP) .  We thank Felix von Oppen, D.\ E.\ Browne, R.\ Gallego, M.\ J.\ Kastoryano and James Wootton for related discussions and comments.

\end{document}